\documentclass[proof]{WileyASNA-v1}

\articletype{Article Type}%

\received{?? September 2022}
\revised{?? ?? 2022}
\accepted{?? ?? 2023}

\begin{document}

\title{On the infrared coincidence: what is the jet contribution to the X-ray power law in GX 339--4?}

\author[1]{David M. Russell*$^{1}$}

\authormark{D. M. Russell}

\address[1]{\orgdiv{Center for Astro, Particle and Planetary Physics}, \orgname{New York University Abu Dhabi}, \orgaddress{\state{Abu Dhabi}, \country{United Arab Emirates}}}

\corres{*\email{dave.russell@nyu.edu}}

\abstract{The hard X-ray power law, prominent in the hard state in black hole X-ray binaries, is generally due to thermal Comptonization in the corona. Optically thin synchrotron emission from compact jets is commonly seen at infrared wavelengths in the hard state. The extent of this spectrum to higher energies remains uncertain. Here, a multi-wavelength study of GX~339--4 is presented. The IR to X-ray spectral index is measured and compared to the X-ray spectral index fitted separately. On some dates in which the jet dominates the IR emission, the X-ray power law and the IR to X-ray power law spectral indices are both in the range $\alpha = -0.7 \pm 0.2$ (where $F_{\nu} \sim \nu^{\alpha}$), i.e. photon index, $\Gamma = 1.7 \pm 0.2$. This suggests they could be the same power law with the same origin, or that this is a coincidence. On other dates in the hard state, $\alpha_{\rm IR-X}<\alpha_{\rm X}$, ruling out a common origin. It is likely that Comptonization dominates on most dates, as expected. However, the X-ray power law never appears to be fainter than the jet power law extrapolated from IR to X-ray, implying that the jet contribution imposes a lower limit to the X-ray flux. If confirmed, this would imply the cooling break in the synchrotron spectrum probably resides at X-ray or higher energies. It is suggested that X-ray spectral fitting should include an extra power law with a break (ideally fit to IR too).}


\maketitle

\section{Introduction}\label{sec1}

\subsection{X-ray binary jets}\label{sec1a}

The radio emission of X-ray binaries (XRBs; in which a black hole or a neutron star is feeding matter from a companion star) is associated with fast outflows of matter in the form of collimated jets~\citep[for a review see][]{FenderGallo2014}. 
Compact jets are relativistic~\citep[e.g.][]{Tetarenko2019,Saikia2019} and are continuously launched during `hard' X-ray states.
The flat/inverted spectrum extends from radio frequencies up to mm or infrared (IR), where it breaks to an optically thin synchrotron spectrum (hereafter referred to as the \textit{jet spectral break}), from particles close to the jet base~\citep[e.g.][ and references therein]{Russell2013}. This optically thin synchrotron emission (OTSE) is emitted from a particle distribution thought to be located at distances of $\sim 10^3$--$\sim 10^5$ gravitational radii ($R_{\rm g}$) from the black hole~\citep[e.g.][]{Markoff2005,Gandhi2017}.

GX~339--4 (4U~1658--48) is a recurrent transient BHXB. In the hard spectral state of GX 339--4 \citep[see e.g.][for definitions of X-ray states]{Belloni2010}, the radio spectral index of the compact jet is inverted ($\alpha \geq 0$, where $F_{\nu}\propto \nu^{\alpha}$) and the far-IR flux agrees with the extrapolation of this power law~\citep{Corbel2013}. The break in the jet spectrum has been identified in the mid-IR~\citep{Gandhi2011}, and the NIR emission is a combination of OTSE from the jet, and thermal emission from the accretion disc~\citep[e.g.][]{CorbelFender2002,CadolleBel2011}. 
The X-ray heated disc (X-ray reprocessing on the disc surface) and the underlying viscous disc are observed at optical to soft X-ray frequencies~\cite[e.g.][]{Homan2005,Tetarenko2020}.

Evidence for the NIR emission to be dominated by OTSE from the compact jet comes 
from its correlation with the X-ray flux, the wavelength-dependent fading and recovery over state transitions, its rapid variability properties including lags with respect to X-rays, and linear polarisation~\citep[e.g.][]{Coriat2009,Casella2010,Russell2011pol,Corbel2013,Vincentelli2019}. 
IR emission from the jet fades over the transition from the hard state to the hard--intermediate state (HIMS), and recovers when the source is fully back in the hard state~\citep{Homan2005,CadolleBel2011,Corbel2013,Kalemci2013}.

\subsection{Could the optically thin synchrotron jet spectrum extend to X-ray energies?}\label{sec1c}

It was first noticed in XTE J1118+480 and GX~339-4, that the optically thin synchrotron jet spectrum seen in the IR, if extrapolated to higher energies, comes close in flux and spectral slope to the observed X-ray power law in the hard state~\citep{Markoff2001,Markoff2003,CorbelFender2002}. Indeed, it was found that the whole broadband spectrum could, in some black hole X-ray binaries \citep[BHXBs; and a few neutron star XRBs; see][]{Lewis2010,Baglio2022}, at some stages in the hard state, potentially be jet dominated~\citep[e.g.][]{Markoff2001,FenderGalloJonker2003,Russell2010}. 
\cite{Hynes2003} found that the rapid variability of XTE J1118+480 has a power law spectrum with $\alpha = -0.6$ from IR to X-ray frequencies, also suggesting a common origin. \cite{Nowak2005} asked the question \textit{``Is the IR coincidence just that?''}, referring to the apparent agreement in GX 339--4 between the extrapolated radio and X-ray power laws, and the IR spectral break.

In addition, some semi-analytical jet models included emission from not just the synchrotron jet from particles downstream of the acceleration and collimation zone (ACZ), but also lepto-hadronic emission, and synchrotron from the jet base (that could contribute primarily to the UV emission), and synchrotron self-Compton emission from the jet base~\cite[that could come from the same particle distribution as in the corona, and contribute to the X-ray flux; e.g.][]{Markoff2005,Connors2019,Kantzas2021}.
In some systems, a jet dominated scenario can be categorically ruled out -- the optically thin synchrotron jet spectrum, extrapolated from the IR, can be a few orders of magnitude fainter than the observed X-ray spectrum at some epochs in the hard state~\citep{Migliari2010,Russell2020}.

In Cyg X--1, jet models were able to explain not the X-ray spectrum, but the high energy $\gamma$-ray tail and reported high level of $\gamma$-ray polarization~\citep{RussellShahbaz2014,Zdziarski2014,Zdziarski2017}. 
A similar model was fitted to the broadband spectrum of GRS~1716--249, with the jet able to account for the radio to mid-IR, and soft $\gamma$-ray spectrum~\citep{Bassi2020,Saikia2022}.
In order for the OTSE power law to extend to MeV energies, models predict that very efficient particle acceleration is required~\citep[e.g.][]{Zdziarski2014,Plotkin2016}. Rapid radiative cooling of electrons produces a break in the synchrotron spectrum (hereafter referred to as the \textit{cooling break}),
which could instead exist in the UV part of the broadband spectrum, requiring less extreme particle acceleration, and implying that the jet contribution to the X-ray spectrum is negligible in most cases. 

\subsection{Evidence for Comptonization}\label{sec1d}

The hard power law that successfully fits the X-ray emission in the hard state, is commonly interpreted as Compton upscattering of soft photons on hot electrons in a corona surrounding the compact object~\citep[e.g.][]{ThornePrice1975}. Such Comptonization models have been tremendously successful in explaining many of the X-ray characteristics~\citep[see e.g.][for a review]{Gilfanov2010}. X-ray spectral models that include an inner accretion disc, a Comptonized corona and a reflection component can very well describe hard state spectra in many BHXBs. 
Comptonization models have been very successful in explaining (i) the power law slope and high energy cut-off often detected at $\sim 100$ keV, (ii) the reflection features, namely the Fe line and Compton hump (reflection of jet synchrotron emission is thought to be diminished if the jet is outflowing and relativistic), and (iii) the Fourier frequency-dependent hard and soft lags~\citep[with the disc producing soft X-rays; e.g.][]{Uttley2011}. 
For example, the high energy cut-off, routinely observed in high luminosity hard states, is very well explained by Compton thermal models, and a jet spectrum is expected to have a much smoother break rather than a sharp cut-off~\citep[e.g.][]{Gilfanov2010,Zdziarski2014}.

Recent, sophisticated models have been able to explain differences in power spectra, and evolution of reverberation lags between hard, intermediate and soft states~\citep[e.g.][]{Wang2022}. Although these models do not favour a jet origin to the X-ray power law, interestingly, some recent models infer properties of the corona that seem to be linked to the jet. Some Comptonization models and reverberation mapping have inferred a change in the size of the corona, or a 2-component corona, or a change in vertical height of the corona~\citep[e.g.][]{Kara2019,Karpouzas2021,Garcia2021}.
Some models even infer an outflowing Comptonizing region or a jet-emitting disc~\citep[e.g.][]{Reig2003,Ferreira2006,Kylafis2012,ReigKylafis2021,You2021}. 
Multi-wavelength studies have also inferred several links between the corona and the 
jet~\citep[e.g.][]{Mendez2022}.
An association between the type B quasi-periodic oscillation (QPO) and the jet has been established from several observations and lines of reasoning~\citep[e.g.][]{MillerJones2012,Russell2020,Homan2020}.
The physical size of the vertical extension of the corona 
can reach up to hundreds of $R_{\rm g}$, which is 
getting towards the distances of the post-accelerated particles in the OTSE region of the jet. It is therefore worth re-visiting the question: ``Can synchrotron emission from the jet contribute to the X-ray spectrum itself?'

\section{The jet contribution to the X-ray power law}\label{sec2}

The spectral index of OTSE from relativistic particles is expected to be $\alpha_{\rm thin} = -0.7 \pm 0.2$~\citep[i.e. a lepton energy distribution of $p = 2.4 \pm 0.4$;][and references therein]{PeerCasella2009,Migliari2010,Russell2013}. Extrapolating the IR power law to higher frequencies, using the measured IR flux of the jet (i.e. removing any contribution from the disc or other components), assuming $\alpha_{\rm thin} = -0.7 \pm 0.2$, one can compare this predicted jet spectrum to both the photon index and the normalization of the observed X-ray power law. 
Considering the following possibilities, one can constrain the synchrotron jet contribution to the X-ray flux.

1. The extrapolated jet spectrum under-predicts the observed X-rays; the X-ray spectrum is brighter than the synchrotron jet. 

2. The extrapolated jet spectrum and the observed X-ray spectrum agree (within errors) both in normalization and spectral index ($\alpha_{\rm X} \sim \alpha_{\rm thin}$). The IR to X-ray spectrum is consistent with one single power law, which could be a coincidence, or it could be from the jet synchrotron emission.

3. The extrapolated jet spectrum over-predicts the X-ray power law, requiring the cooling break to reside in the optical--UV, whether the jet contributes to the X-ray flux or not.

4. The jet spectrum agrees with the normalization of the X-ray spectrum, but the observed X-ray photon index disagrees with the IR to X-ray spectral index ($\alpha_{\rm X} \neq \alpha_{\rm thin}$), so the synchrotron jet cannot account for the X-ray spectrum.

\section{Infrared and X-ray data}\label{sec3}

GX 339--4 has a large wealth of multi-wavelength monitoring during many outbursts. For this initial study, we take X-ray data from~\cite{Dunn2010} and IR data from~\cite{Buxton2012,Dincer2012}. The X-ray data are \textit{Rossi X-ray Timing Explorer (RXTE)} observations. The \textit{Proportional Counter Array (PCA)} and \textit{High Energy X-ray Timing Experiment (HEXTE)} spectra are fitted together with a model that includes a multi-temperature accretion disc, a (broken) power law, an iron line at 6.4 keV, a high energy break, and absorption from neutral hydrogen~\citep{Dunn2010}. The IR photometric data are from the \textit{Small \& Moderate Aperture Research Telescope System}~\citep[\textit{SMARTS}; using the \textit{CTIO} 1.3~m telescope; see][for details]{Buxton2012}.

In order to estimate the IR jet flux, we take the $H$-band magnitudes (central wavelength 1.6 $\mu$m; the longest wavelength filter, which has the highest relative jet contribution) and de-redden them using the extinction $A_{\rm V} = 3.25$~\citep{Gandhi2011} and $A_{\rm H} = 0.19 A_{\rm V}$~\citep{Cardelli1989}. 
We compare the measured IR to X-ray spectral index, $\alpha_{\rm IR-X}$ (measured directly from the IR and X-ray power law flux measurements), to the X-ray spectral index measured independently by~\cite{Dunn2010}.
From analysis of the $I-H$ magnitude colours of all the \textit{SMARTS} observations, we find that the jet dominates the IR flux in the hard state at magnitudes $H\leq 14.5$. At fainter magnitudes, the $I-H$ colour becomes similar to the soft state data, in which the disc dominates. For data in the soft state, and any data fainter than $H=14.5$, we cannot extrapolate the jet spectrum to higher frequencies because the IR is likely to contain a prominent disc component.

\begin{figure}[t]
\centerline{\includegraphics[angle=0,width=8.5cm]{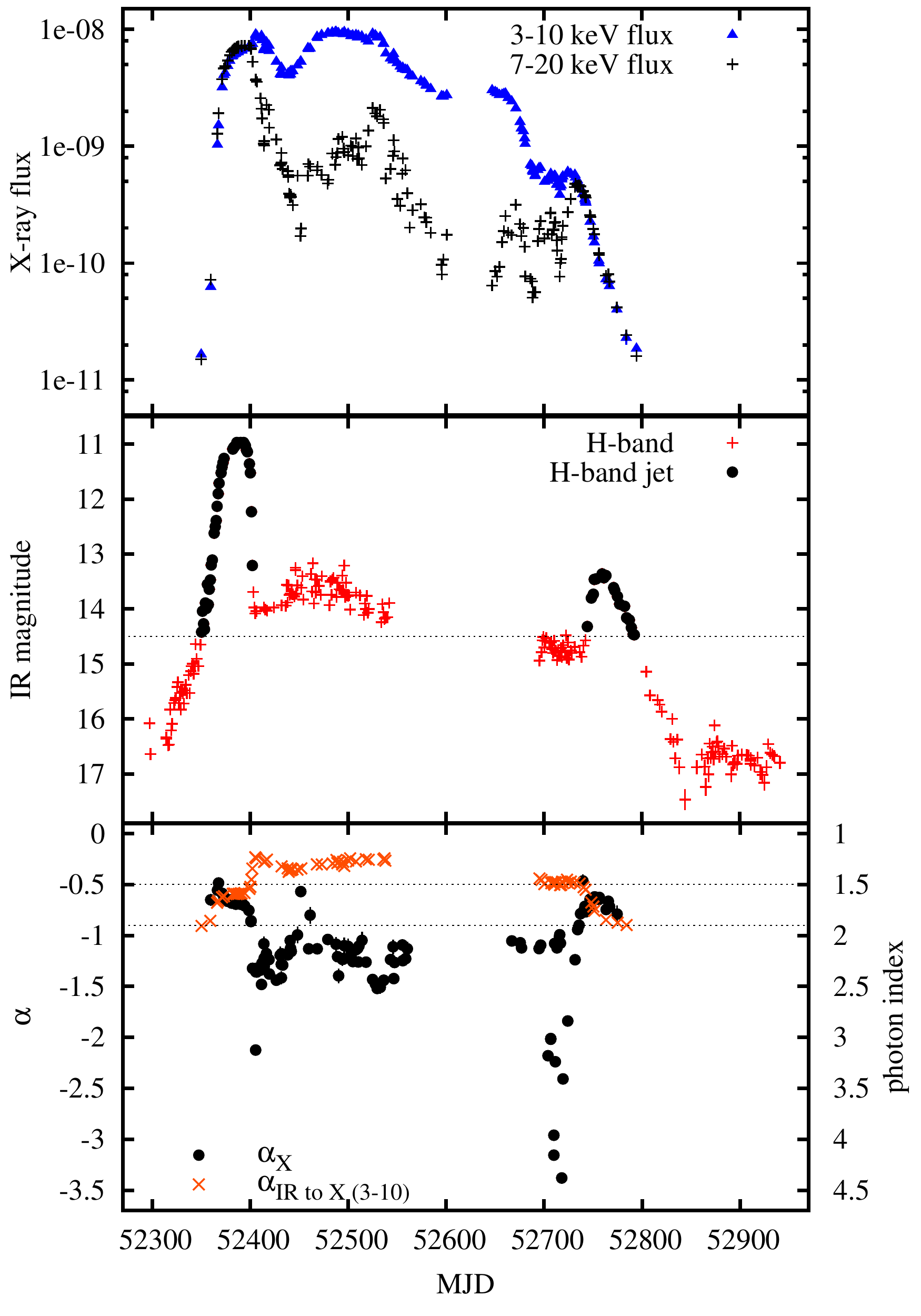}}
\caption{IR and X-ray light curves of the 2002--2003 outburst of GX 339--4. \textit{Upper panel:} soft X-ray (3--10 keV; blue triangles) and hard X-ray (7--20 keV; black pluses) fluxes. \textit{Middle panel:} Observed IR magnitude (red pluses), with jet-dominated points (black circles); the dashed horizontal line denotes $H=14.5$ mag (see text for details). \textit{Lower panel:} the X-ray spectral index, $\alpha_{\rm X}$ (black circles; photon index $\Gamma$ on the right axis), and the IR to X-ray spectral index, $\alpha_{\rm IR-X}$ (orange crosses). The two dashed horizontal lines indicate the range of spectral indices expected for OTSE ($\alpha_{\rm thin} = -0.7 \pm 0.2$). \label{lc2002}}
\end{figure}

\begin{figure}[t]
\centerline{\includegraphics[angle=0,width=8.5cm]{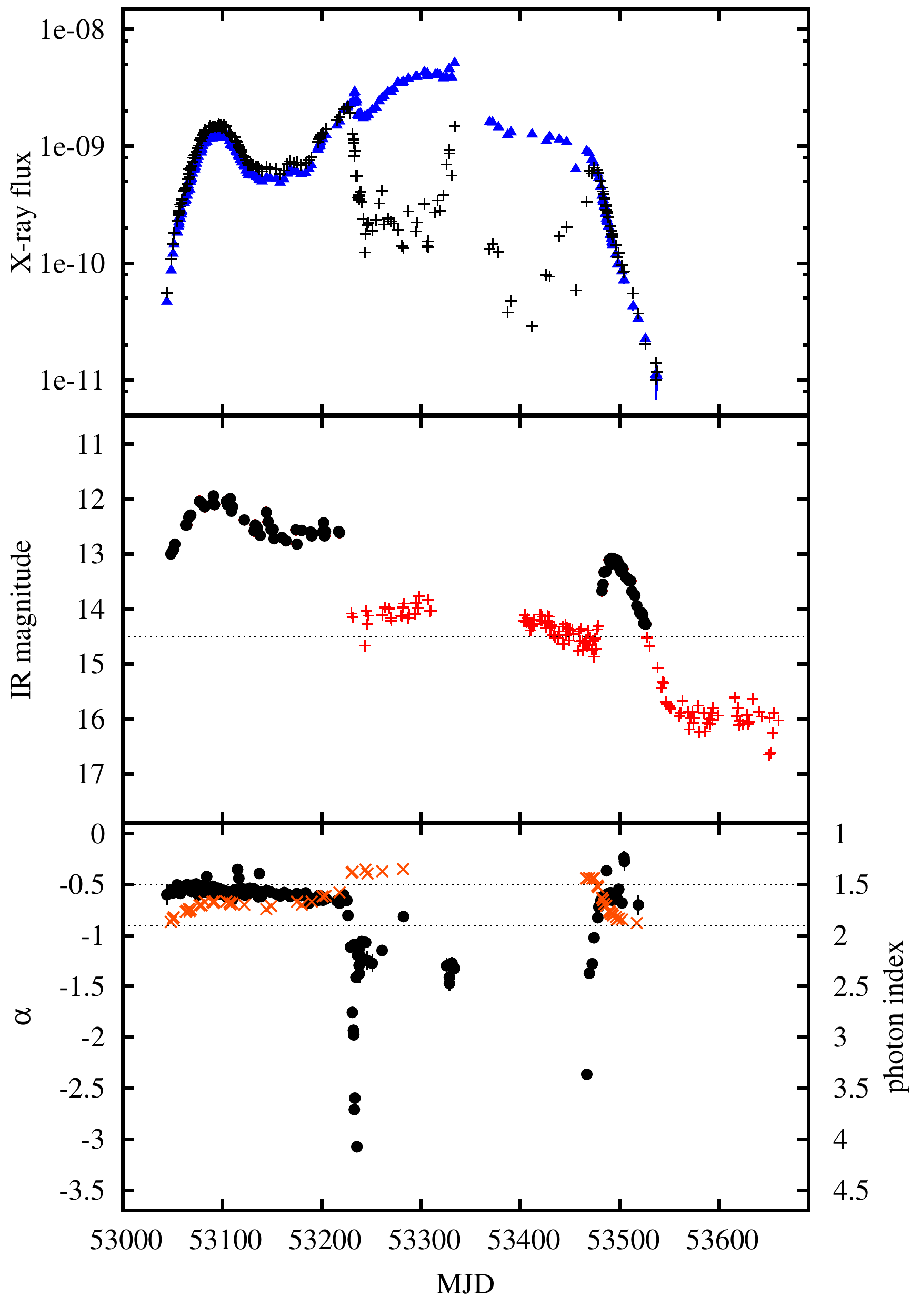}}
\caption{IR and X-ray light curves of the 2004--2005 outburst (see text for details). Symbols are the same as Fig.~\ref{lc2002}.\label{lc2004}}
\end{figure}

\begin{figure}[t]
\centerline{\includegraphics[angle=0,width=8.5cm]{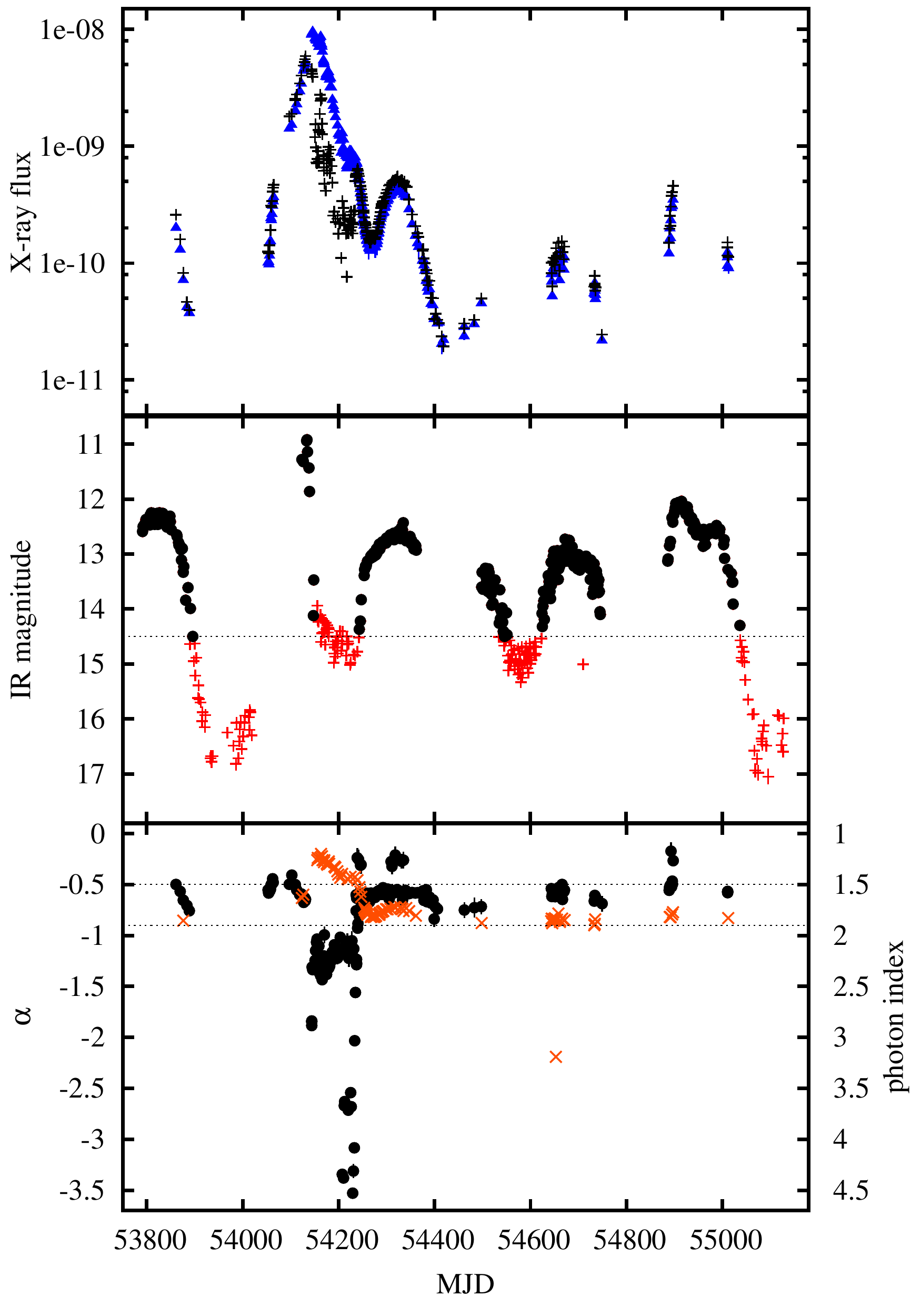}}
\caption{IR and X-ray light curves of the 2006--2009 activity period (see text). Symbols are the same as Fig.~\ref{lc2002}.\label{lc2006}}
\end{figure}

\section{Results and Analysis}\label{sec4}

\subsection{Light curves}\label{sec4a}

In Fig.~\ref{lc2002}, the X-ray and IR light curves of the 2002--2003 outburst of GX 339--4 are shown in the upper and middle panels, respectively. 
Both the soft (3--10 keV; blue triangles) and hard (7--20 keV; black pluses) X-ray fluxes rise together at the start of the outburst, then they separate over the transition, with the hard X-ray flux fading and the soft X-ray flux remaining bright. 
During the return transition to the hard state,
the hard X-ray flux brightened then the two faded together. The IR flux brightened at the start of the outburst, before fading dramatically over the state transition. 
The IR flux brightened on return to the hard state, then faded during decay towards quiescence. In Fig.~\ref{lc2002} (middle panel), the black circles are the data points in which we are confident the jet dominates the IR emission (see section~\ref{sec3}). The red pluses are data fainter than this magnitude, and also data in the soft state, when the IR jet emission cannot be measured.
In the lower panel of Fig.~\ref{lc2002}, the X-ray spectral index, $\alpha_{\rm X}$ 
is shown as black circles, and the IR to X-ray spectral index
($\alpha_{\rm IR-X}$), is shown as orange crosses. 
X-ray data which had 
spectral index uncertainties greater than 0.1, were excluded from the following analysis, since accurate spectral index values are required to compare to the IR to X-ray spectral index. The IR to X-ray spectral index errors are very small.
In Figs.~\ref{lc2004} and~\ref{lc2006}, the X-ray and IR light curves of the 2004--2005 outburst, and the 2006--2009 period of activity, are presented (panels and symbols are the same as Fig.~\ref{lc2002}).

\subsection{Comparing spectral indices}\label{sec4b}

From the lower panels of Figs.~\ref{lc2002}--\ref{lc2006}, it is evident that during the hard state, on some dates the IR to X-ray spectral index and the X-ray spectral index agree ($\alpha_{\rm IR-X} \sim \alpha_{\rm X}$). However on other dates, they do not, with the X-ray spectral index being shallower than the IR to X-ray spectral index ($\alpha_{\rm X} > \alpha_{\rm IR-X}$), or occasionally the opposite ($\alpha_{\rm X} < \alpha_{\rm IR-X}$). Over the transitions and in the soft state, there are large differences between the two values.
During 2006--2009, the source had a full outburst with state transitions around MJD~54000--54400 (2006--2007), and was active in the hard state before and after this (in early 2006, and in late 2007 to 2009). This hard state activity could be classified as mini-outbursts or failed transition (FT) outbursts~\cite[e.g.][and references therein]{Alabarta2021}. 
During the period of hard state activity (FT outbursts), the difference in the spectral indices is quite striking, with $\alpha_{\rm X} \sim -0.6$ and $\alpha_{\rm IR-X}\sim -0.9$.
The spectral indices do not agree within errors, for the majority of data in this time period (Fig.~\ref{lc2004}).

\begin{figure}[t]
\centerline{\includegraphics[angle=270,width=9cm]{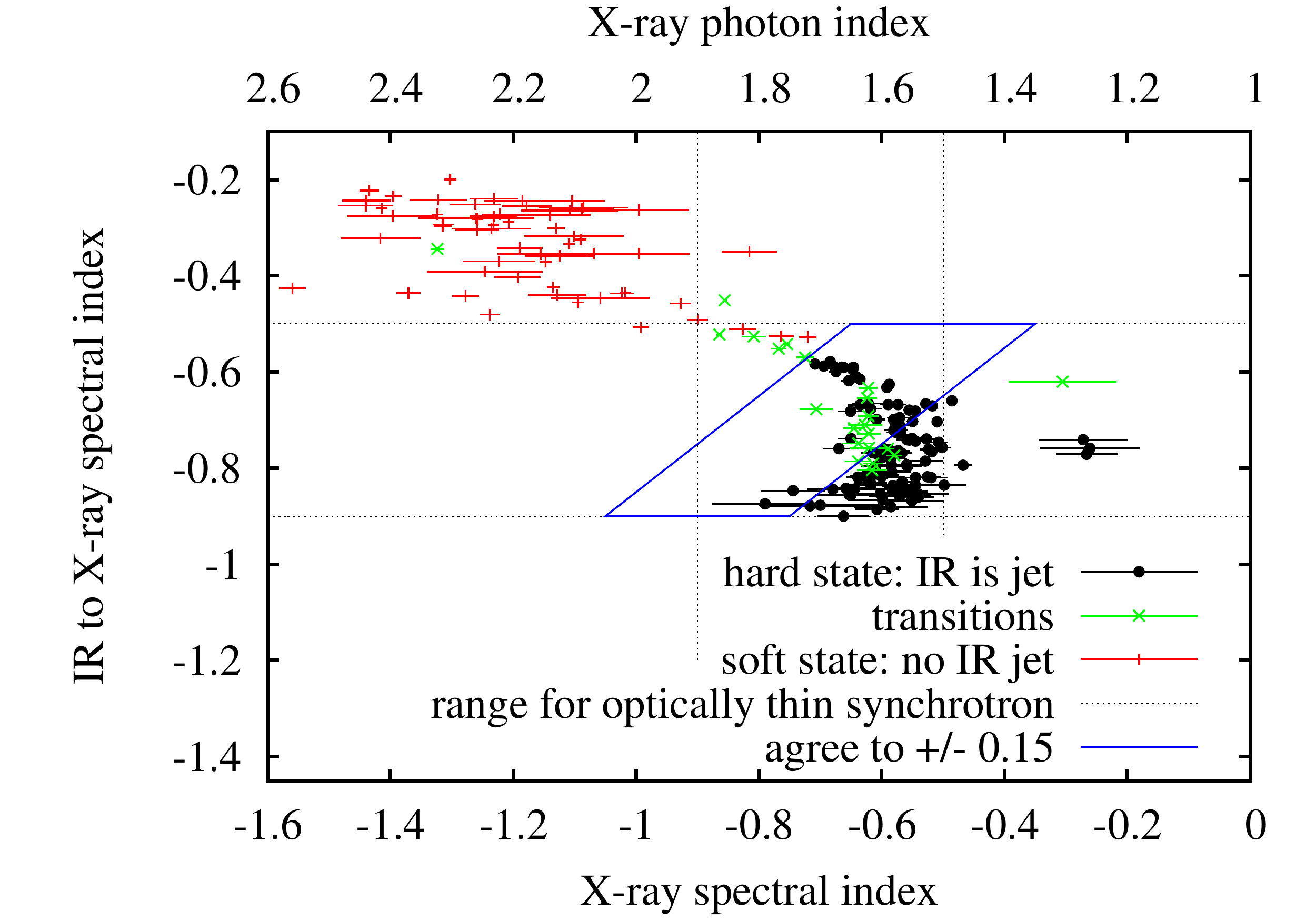}}
\caption{The X-ray spectral index $\alpha_{\rm X}$, vs the IR to X-ray spectral index $\alpha_{\rm IR-X}$ (see text for details).\label{alphaalpha}}
\end{figure}

In Fig.~\ref{alphaalpha}, the X-ray spectral index $\alpha_{\rm X}$, vs the IR to X-ray spectral index $\alpha_{\rm IR-X}$ is plotted (data taken within 24 hr).
The X-ray photon index is indicated on the top axis. The two vertical dotted lines indicate the range $-0.9 < \alpha_{\rm X} < -0.5$ and the two horizontal dotted lines show the range $-0.9 < \alpha_{\rm IR-X} < -0.5$. Any data that exist within the box created by these lines, are consistent with the spectral index expected for the jet, from IR to X-ray. The region encompassed by the blue parallelogram shows data with $-0.9 < \alpha_{\rm IR-X} < -0.5$, which also have the two spectral indices agreeing within $\pm 0.15$.

Fig.~\ref{alphaalpha} shows that almost all of the data in the soft state 
(red pluses) 
have $\alpha_{\rm IR-X} > -0.5$ and $\alpha_{\rm X} < -0.9$ (photon index $\Gamma > 1.9$), both of which are inconsistent with jet emission. 
Almost all of the data in the hard state (black circles) and transition states (green crosses) however, reside within the $-0.9 < \alpha < -0.5$ box, so both $\alpha_{\rm X}$ and $\alpha_{\rm IR-X}$ are consistent with OTSE. However, a large fraction of the hard state data points lie outside the blue region, having $\alpha_{\rm X} > \alpha_{\rm IR-X}$. For these points, the two spectral indices disagree, so the X-ray power law is likely produced by Comptonization in the corona, not OTSE from the jet. For the data points inside the blue region, we cannot rule out the jet dominating the entire IR to X-ray spectrum. The data here are consistent with a single power law with spectral index -0.7 to -0.6, typical of OTSE. However, many data points have small enough spectral index errors to state that the two values are not consistent with each other, within the errors.
What is intriguing is that there are no data in the lower part of the plot. All data have $\alpha_{\rm IR-X} > -0.9$. This implies that the observed X-ray flux is never fainter than the extrapolated jet spectrum. This is useful, as it means that a lower limit to the X-ray flux can be deduced from the IR jet flux. The implications -- and this is model-independent, is that the extrapolated jet flux imposes a lower limit to X-ray flux. This result is remarkable, because it suggests that maybe the jet does contribute to the X-ray flux -- when the corona becomes fainter, the jet is revealed. If confirmed, it would also suggest that the cooling break in the jet spectrum is at X-ray or higher frequencies.

In a follow-up work, differences will be searched for in the X-ray spectral and timing properties, and reflection component properties, as a function of position on the $\alpha_{\rm X}$ -- $\alpha_{\rm IR-X}$ diagram.
If there is a possibility that the jet power law could contribute significantly, then it should be included in X-ray fitting models. An extra power law with a curved break (ideally fit to IR too) could be included in spectral fits of hard state BHXBs. It is possible that the jet power law (with a break at $\sim$ tens of keV) could be ‘masquerading’ as a feature commonly fitted as a Compton hump from reflection. Such additional component fitting is already being included in some cases~\citep[e.g.][]{Zdziarski2014,Bassi2020}.

\section*{Acknowledgments}

This paper has made use of \textit{SMARTS} optical and infrared data. I thank Robert Dunn for providing the \textit{RXTE} data analysis from~\cite{Dunn2010}, and several experts for very fruitful discussions on this topic, including Sera Markoff, Matteo Lucchini, Chris Done, Phil Uttley, Mariano M{\'e}ndez, Cristina Baglio, Kevin Alabarta and Pierre-Olivier Petrucci.

\bibliography{jet-Xray-XMM-AN}

\end{document}